\documentclass[12pt,reqno,fleqn]{amsart}

\usepackage{latexsym}
\usepackage{amssymb}
\usepackage{amsxtra}
\usepackage{amsmath}
\usepackage{amsfonts}
\usepackage{CJK}
\newcommand{\cleqn}{\setcounter{equation}{0}}
\newcommand{\clth}{\setcounter{theorem}{0}}
\newcommand {\sectionnew}[1]{\section{#1}\cleqn\clth}
\newtheorem{theorem}{Theorem}[section]
\newtheorem{lemma}[theorem]{Lemma}

\newtheorem{remark}[theorem]{Remark}
\def\({\left(}
\def\){\right)}
\def\[{\left[}
\def\]{\right]}
\def\d{\partial}

\begin{document}

\title{Quantum Torus symmetries of the CKP and multi-component CKP hierarchies}
\author{Qiufang Liu, Chuanzhong Li$^*$}
\dedicatory {  Department of Mathematics, Ningbo University, Ningbo, 315211 Zhejiang, P.\ R.\ China}

\thanks{$^*$ Corresponding author: lichuanzhong@nbu.edu.cn}

\texttt{}
\date{}

%%%%%%%%%%%%%%%%%%%%%%%%%%%%%%%%%%%%%%%%%%%%%%%%
\begin{abstract}
In this paper, we construct a series of additional flows of the CKP hierarchy and the multi-component CKP hierarchy  and these flows constitute a \emph{N}-folds direct product of the positive half of the quantum Torus symmetry. Comparing to the $\emph{W}_{\infty}$ infinite dimensional Lie symmetry, this quantum Torus symmetry has a nice algebraic structure with double indices.
\end{abstract}

\maketitle
%\tableofcontents
\allowdisplaybreaks
 \setcounter{section}{0}

\sectionnew{Introduction}

The KP hierarchy is one of the most important integrable hierarchies and it arises in many different fields of mathematics and physics such as enumerative algebraic geometry, topological field and string theory. It has perfect structures such as the Virasoro type additional symmetry which has been extensively studied in literature(\cite{B},\cite{C},\cite{A},\cite{D},\cite{E}). Date, Jimbo, Kashiwara and Miwa extended their work on the KP hierarchy to the multi-component KP hierarchy in(\cite{T}). The additional symmetries have been studied by Orlov and Schulman about the explicit form of the additional flows of the KP hierarchy in(\cite{C}). The dynamic variables are included in this kind of additional flows which form a centerless $W_{1+\infty}$ algebra. However, by using the Virasoro constraint and string equations, this algebra is closely related to the theory of matrix models(\cite{F},\cite{G}). In the literatures(\cite{H},\cite{I},\cite{J},\cite{K}), the Block algebra which is a generalization of Virasoro algebras has been studied extensively in the field of Lie algebras. In addition, Block type additional symmetries of the bigraded Toda hierarchy have been given in the paper(\cite{L}). Later, a lot of studies on integrable systems and Block algebras have been shown, such as(\cite{M},\cite{N},\cite{O}). Using the quantization, the Block Lie algebra can be generalized to the quantum torus Lie algebra that is shown out in a few literatures(\cite{P},\cite{Q}).\\

As we all know, the BKP hierarchy and the CKP hierarchy are two sub-hierarchies  of the KP hierarchy. However, many studies on additional symmetries about the BKP hierarchy have been done, such as the additional symmetry of the BKP hierarchy(\cite{S}), the quantum torus symmetry of the BKP hierarchy(\cite{Q}), the additional symmetry of the supersymmetric BKP hierachy (\cite{R})  and so on. About the CKP hierarchy, the additional symmetry of the CKP hierarchy has been given in the paper(\cite{A}), with the above preparation, we should pay our attention on the quantum torus symmetry of  the CKP hierarchy and the multi-component CKP hierarchy. \\

The organization of this paper is as follows. We firstly review the Lax equation of the CKP hierarchy in section 2. In section 3, under the basic Sato theory, we construct the additional symmetry of the CKP hierarchy which forms a  quantum torus Lie algebra. We will give a brief description of the multi-component CKP hierarchy in section 4. In section 5, the quantum torus symmetry is further generalized to the multi-component CKP hierarchy.

\sectionnew{The CKP hierarchy}
The pseudo-differential operator is  as
\begin{equation}
L=\d+u_1\d^{-1}+u_2\d^{-2}+u_3\d^{-3}+...,
\end{equation}
and then the KP hierarchy is defined by
\begin{equation}
\frac{\d L}{\d t_n}=[B_n,L], n=1,2,3,....
\end{equation}
Here $B_n=(L^{n})_{+}=\sum\limits_{k=0}^n a_k\d^{k}$ denotes the non-negative powers of $\d$ in $L^{n}$, $\d=\d/\d x$, $u_{i}=u_{i}(x=t_1,t_2,t_3,...,).$ In order to define the CKP hierarchy, we need a formal adjoint operation $*$ for an arbitrary pseudo-differential operator $P=\sum\limits_{i}p_{i}\d^{i}$, here we have $P^{*}=\sum\limits_{i}(-1)^{i}\d^{i}p_{i}.$ For example, $\d^{*}=-\d$, $(\d^{-1})^{*}=-\d^{-1}$, and $(AB)^{*}=B^{*}A^{*}$ for two operators $A,B$. The CKP hierarchy is a reduction of the KP hierarchy by the constraint
\begin{equation}
L^{*}=-L
\end{equation}
which freeze all even flows of the KP hierarchy, i.e. the Lax equation of the CKP hierarchy has only odd flows,
\begin{equation}
\frac{\d L}{\d t_{2n+1}}=[B_{2n+1},L], n=0,1,2,...,
\end{equation}
thus $u_{i}=u_{i}(t_1,t_3,...)$ for the CKP hierarchy.\\
Basing on the definition, the Lax operator of the CKP hierarchy has form
\begin{equation}
\mathcal{L}_{c}=\d+\sum\limits_{i\geq1}u_{i}\d^{-i},
\end{equation}
such that
\begin{equation}\label{a}
\mathcal{L}_{c}^{*}=-\mathcal{L}_{c}.
\end{equation}
We call eq.(\ref{a}) the $C$ type condition of the CKP hierarchy.
The CKP hierarchy is defined by the following Lax equations:
\begin{equation}
\frac{\d \mathcal{L}_{c}}{\d t_{k}}=[(\mathcal{L}_{c}^{k})_{+},\mathcal{L}_{c}], k\in \mathbb{Z}_{+}^{odd}.
\end{equation}
However, the operator $\mathcal{L}_{c}$ can be generated by a dressing operator $\Phi_{c}=1+\sum\limits_{j=1}^{\infty}\omega_{j}\d^{-j}$ in the following way
\begin{equation}
\mathcal{L}_{c}=\Phi_{c}\d\Phi_{c}^{-1},
\end{equation}
in which $\Phi_{c}$ satisfies
\begin{equation}
\Phi_{c}^{*}=\Phi_{c}^{-1}.
\end{equation}
At the same time, the dressing operator $\Phi_{c}$ needs to satisfy the following Sato equation
\begin{equation}\label{e}
\frac{\d \Phi_{c}}{\d t_{k}}=-(\mathcal{L}_{c}^{k})_{-}\Phi_{c}, k=1,3,5,...
\end{equation}

\sectionnew{Quantum torus symmetry of the CKP hierarchy}
Basing on the above dressing structure and Sato equations, it is convenient that we construct the Orlov-Schulman's operator which is used to give the quantum torus symmetry of the CKP hierarchy. In the next section, the additional symmetry and  the algebraic structure of the additional flows of the CKP hierarchy will be discussed.\\
At first, we define the operator $\Gamma_{c}$ and the Orlov-Schulman's operator $\mathcal{M}_{c}$ as
\begin{equation}
\Gamma_{c}=\sum\limits_{i\in \mathbb{Z}_{+}^{odd}}it_i\d^{i-1}, \mathcal{M}_{c}=\Phi_{c}\Gamma_{c}\Phi_{c}^{-1}.
\end{equation}
The Lax operator $\mathcal{L}_{c}$ and the Orlov-Schulman's operator $\mathcal{M}_{c}$ satisfy the following canonical relation
\begin{equation}
[\mathcal{L}_{c},\mathcal{M}_{c}]=1.
\end{equation}
The Lax equation of the CKP hierarchy can be given by the consistent conditions of the following set of linear partial differential equations
\begin{equation}
\mathcal{L}_{c}\omega_{c}(t;z)=z\omega_{c}(t;z), \frac{\d \omega_{c}(t;z)}{\d t_{2n+1}}=B_{2n+1}\omega_c(t;z), t=(t_1, t_3, t_5,...).
\end{equation}
Here $\omega_{c}(t;z)$ is identified as a wave function, using the wave operator $\Phi_{c}=1+\sum\limits_{j=1}^{\infty}\omega_{j}\d^{-j}$ , the wave function admit the following representation
\begin{equation}
\omega_{c}(t;z)=\Phi_{c}e^{\xi_{c}(t;z)},
\end{equation}
where the function of $\xi_{c}$ is defined as $\xi_{c}(t;z)=\sum\limits_{k\in \mathbb{Z}_{+}^{odd}}t_{k}z^{k}$.
With the above preparation, it is time to construct additional symmetries for the CKP hierarchy in the next section. Basing on the above basic knowledge, we can get that the operator $\mathcal{M}_{c}$ satisfy
\begin{equation}
[\mathcal{L}_{c},\mathcal{M}_{c}]=1, \mathcal{M}_{c}\omega_{c}(z)=\d_{z}\omega_{c}(z);
\end{equation}
\begin{equation}
\frac{\d \mathcal{M}_{c}}{\d t_{k}}=[(\mathcal{L}_{c}^{k})_{+},\mathcal{M}_{c}], k\in \mathbb{Z}_{+}^{odd}.
\end{equation}
Given any pair of integers $(m,n)$, where $m\geq0, n\geq0$, we will have the following monomials in $\mathcal{L}_{c}$ and $\mathcal{M}_{c}$
\begin{equation}\label{b}
A_{mn}=\mathcal{M}_{c}^{m}\mathcal{L}_{c}^{n}-(-1)^{n}\mathcal{L}_{c}^{n}\mathcal{M}_{c}^{m}.
\end{equation}
For any monomials $A_{mn}$ in $(\ref{b})$, we can get
\begin{equation}
\frac{\d A_{mn}}{\d t_{k}}=[(\mathcal{L}_{c}^{k})_{+},A_{mn}], k\in \mathbb{Z}_{+}^{odd}.
\end{equation}
Next, we should prove $A_{mn}$ satisfy a $C$ type condition with the following lemma.
\begin{lemma}\label{c}
The operator $\mathcal{M}_{c}$ satisfies the following identity
\begin{equation}
\mathcal{M}_{c}^{*}=\mathcal{M}_{c}.
\end{equation}
\end{lemma}
\textbf{Proof}. By using
\begin{equation}
\Phi_{c}^{*}=\Phi_{c}^{-1}, \Gamma_{c}^{*}=\Gamma_{c},
\end{equation}
then we have
\begin{equation}
\mathcal{M}_{c}^{*}=(\Phi_{c}\Gamma_{c}\Phi_{c}^{-1})^{*}=(\Phi_{c}^{-1})^{*}\Gamma_{c}^{*}\Phi_{c}^{^{*}}=\Phi_{c}\Gamma_{c}\Phi_{c}^{-1}=\mathcal{M}_{c}.
\end{equation}
Using the lemma $(\ref{c})$ above, it is easy to check that the operator $A_{mn}$ satisfy the $C$ type condition, we have
\begin{equation}\label{d}
A_{mn}^{*}=-A_{mn}.
\end{equation}
Now we will denote the operator $C_{mn}$ as
\begin{equation}
C_{mn}:=e^{m\mathcal{M}_{c}}q^{n\mathcal{L}_{c}}-q^{-n\mathcal{L}_{c}}e^{m\mathcal{M}_{c}},
\end{equation}
which will further leads to
\begin{equation}\label{g}
C_{mn}=\sum\limits_{p,s=0}^{\infty}\frac{m^p(n\log q)^s(\mathcal{M}_{c}^{p}\mathcal{L}_{c}^{s}-(-1)^s\mathcal{L}_{c}^{s}\mathcal{M}_{c}^{p})}{p!s!}=\sum\limits_{p,s=0}^{\infty}\frac{m^p(n\log q)^sA_{ps}}{p!s!}.
\end{equation}
By using the eq.(\ref{d}), we can get the $C$ type anti-symmetry property of $C_{mn}$ by the following calculation
\begin{eqnarray*}
C_{mn}^{*}&=&(\sum\limits_{p,s=0}^{\infty}\frac{m^p(n\log q)^sA_{ps}}{p!s!})^{*}\\
&=&-\sum\limits_{p,s=0}^{\infty}\frac{m^p(n\log q)^sA_{ps}}{p!s!}\\
&=&-C_{mn}.
\end{eqnarray*}
Therefore, the operator $C_{mn}$ satisfy the following $C$ type condition
\begin{equation}
C_{mn}^{*}=-C_{mn}.
\end{equation}
On the basis of the quantum parameter $q$, the additional symmetry flows for the time variable $t_{m,n}$, $t_{m,n}^{*}$ are defined as follows
\begin{equation}\label{f}
\frac{\d \Phi_{c}}{\d t_{m,n}}=-(A_{mn})_{-}\Phi_{c}, \frac{\d \Phi_{c}}{\d t_{m,n}^{*}}=-(C_{mn})_{-}\Phi_{c},
\end{equation}
which are equivalent to
\begin{equation}
\frac{\d \mathcal{L}_{c}}{\d t_{m,n}}=-[(A_{mn})_{-},\mathcal{L}_{c}], \frac{\d \mathcal{M}_{c}}{\d t_{m,n}^{*}}=-[(C_{mn})_{-},\mathcal{M}_{c}].
\end{equation}
Similarly, we can also get
\begin{equation}
\d_{t_{l,k}^{*}}(C_{mn})=[-(C_{lk})_{-},C_{mn}].
\end{equation}
\begin{remark}
From the specific construction of the operator $C_{mn}$, we can see that the reduction condition in eq.(\ref{a}) bring important impacts on the generators of the additional symmetry flows.
\end{remark}
The commutativity of the additional flow $\frac{\d}{\d t_{mn}^{*}}$ with the additional flow $\frac{\d}{\d t_{k}}$ can be derived by the following theorem.
\begin{theorem}\label{k}
The additional flows $\d_{t_{m,n}^{*}}$ are symmetries of the CKP hierarchy, which commute with all flows $\d_{t_{k}}$ of the CKP hierarchy, i.e.
\begin{equation}
[\d_{t_{m,n}^{*}},\d_{t_{k}}]=0.
\end{equation}
\end{theorem}
\textbf{Proof}.
Using the eq.(\ref{e}) and eq.(\ref{f}), we have
\begin{eqnarray*}
[\d_{t_{m,n}^{*}},\d_{t_{k}}]\Phi_{c}&=&-\d_{t_{m,n}^{*}}({\mathcal{L}_c}^{k}_{-}\Phi_{c})+\d_{t_{k}}((C_{mn})_{-}\Phi_{c})\\
&=&[(C_{mn})_{-},\mathcal{L}_{c}^{k}]_{-}\Phi_{c}+{\mathcal{L}_c}^{k}_{-}(C_{mn})_{-}\Phi_{c}+[{\mathcal{L}_c}_{+}^{k},C_{mn}]_{-}\Phi_{c}-(C_{mn})_{-}{\mathcal{L}_c}^{k}_{-}\Phi_{c}\\
&=&[(C_{mn})_{-},\mathcal{L}_{c}^{k}]_{-}\Phi_{c}+[{\mathcal{L}_c}_{+}^{k},(C_{mn})_{-}]_{-}\Phi_{c}+[{\mathcal{L}_c}^{k}_{-},(C_{mn})_{-}]\Phi_{c}\\
&=&[(C_{mn})_{-},{\mathcal{L}_c}^{k}_{-}]_{-}\Phi_{c}+[{\mathcal{L}_c}^{k}_{-},(C_{mn})_{-}]\Phi_{c}\\
&=&0
\end{eqnarray*}
As we all know, the additional flows $\d_{t_{l,k}}$ of the CKP hierarchy form the following $W_{\infty}$ algebra which has been shown in \cite{A}
\begin{eqnarray*}
[\d_{t_{p,s}},\d_{t_{a,b}}]\mathcal{L}_{c}=\sum\limits_{\alpha\beta}C_{\alpha\beta}^{(ps)(ab)}\d_{t_{\alpha\beta}}\mathcal{L}_{c}.
\end{eqnarray*}
Now it is  time for us to give the algebraic structure of the additional flows $\d_{t_{l,k}^{*}}$ of the CKP hierarchy.
\begin{theorem}
The additional flows $\d_{t_{l,k}^{*}}$ of the CKP hierarchy form the positive half of quantum torus algebra, we have
\begin{equation}
[\d_{t_{n,m}^{*}},\d_{t_{l,k}^{*}}]=(q^{ml}-q^{nk})\d_{t_{n+l,m+k}^{*}}, n, m, l, k \geq0.
\end{equation}
\end{theorem}
\textbf{Proof}. Similarly to eq.(\ref{g}), we can rewrite the quantum torus flow which is a combination of the additional flows $\d_{t_{m,n}}$ as follows
\begin{eqnarray*}
[\d_{t_{n,m}^{*}},\d_{t_{l,k}^{*}}]\mathcal{L}_{c}
&=&[\sum\limits_{p,s=0}^{\infty}\frac{n^p(m\log q)^{s}}{p!s!}\d_{t_{p,s}},\sum\limits_{a,b=0}^{\infty}\frac{l^a(k\log q)^{b}}{a!b!}\d_{t_{a,b}}]\mathcal{L}_{c}\\
&=&\sum\limits_{p,s=0}^{\infty}\sum\limits_{a,b=0}^{\infty}\frac{n^p(m\log q)^{s}}{p!s!}\frac{l^a(k\log q)^{b}}{a!b!}[\d_{t_{p,s}},\d_{t_{a,b}}]\mathcal{L}_{c}\\
&=&\sum\limits_{p,s=0}^{\infty}\sum\limits_{a,b=0}^{\infty}\frac{n^p(m\log q)^{s}}{p!s!}\frac{l^a(k\log q)^{b}}{a!b!}\sum\limits_{\alpha\beta}C_{\alpha\beta}^{(ps)(ab)}\d_{t_{\alpha\beta}}\mathcal{L}_{c}\\
&=&(q^{ml}-q^{nk})\sum\limits_{\alpha,\beta=0}^{\infty}\frac{(n+l)^{\alpha}((m+k)\log q)^{\beta}}{\alpha!\beta!}\d_{t_{\alpha,\beta}}\mathcal{L}_{c}\\
&=&(q^{ml}-q^{nk})\d_{t_{n+l,m+k}^{*}}\mathcal{L}_{c}.
\end{eqnarray*}
Therefore, we find that the additional flows $\d_{t_{l,k}^{*}}$ constitute a perfect quantum torus algebra.

\sectionnew{The multi-component CKP hierarchy}
In this section, we consider the $\emph{N}$-component CKP hierarchy. There are $\emph{N}$ infinite families of time variables $t_{\alpha,k}$ for it, where $\alpha=1, 2, ..., \emph{N}, k=1, 3, 5, ...$. \\
The Lax operator of the $\emph{N}$-component CKP hierarchy has the form
\begin{equation}
 L_{c}=D\d+u_1\d^{-1}+u_2\d^{-2}+...,
\end{equation}
in which the coefficients $D, u_1, u_2, ...$ are $\emph{N}\times \emph{N}$ matrices and $D=diag(d_1, d_2, ..., d_\emph{N})$.\\
There are another $\emph{N}$ pseudo-differential operators $R_1, R_2, ..., R_\emph{N}$ which are defined as
\begin{eqnarray*}
R_\alpha=E_{\alpha}+u_{\alpha,1}\d^{-1}+u_{\alpha,2}\d^{-2}+...,
\end{eqnarray*}
where $u_{\alpha,1}, u_{\alpha,2}, ...$ are $\emph{N}\times \emph{N}$ matrices, and $E_{\alpha}$ is also $\emph{N}\times \emph{N}$ matrix with $1$ on the $(\alpha,\alpha)$-component and zero elsewhere. The operators $ L_{c}, R_1, ..., R_\emph{N}$ satisfy conditions as follows
\begin{eqnarray*}
 L_{c}R_\alpha=R_\alpha L_{c},\ \   R_\alpha R_\beta=\delta_{\alpha\beta}R_\alpha, \ \  \sum\limits_{\alpha=1}^{\emph{N}}R_\alpha=1.
\end{eqnarray*}
We need a formal adjoint operation $*$ for an arbitrary matrix-valued pseudo-differential operator $P=\sum\limits_{i}p_{i}\d^{i}$, here we have $P^{*}=\sum\limits_{i}(-1)^{i}\d^{i}p_{i}^{T}$, in another, there exists $(AB)^{*}=B^{*}A^{*}$ for two matrix-valued pseudo-differential operators.
$ L_{c}, R_\alpha$ satisfy
\begin{equation}\label{h}
 L_{c}^{*}=- L_{c}, \ \ R_\alpha^*=R_\alpha.
\end{equation}
The Lax equations can be defined as
\begin{equation}
\frac{\d  L_{c}}{\d t_{k}^{\alpha}}=[A_{\alpha,k}, L_{c}],\ \ \frac{\d R_{\beta}}{\d t_{k}^{\alpha}}=[A_{\alpha,k},R_{\beta}],\ \ A_{\alpha,k}:=(L_{c}^{k}R_{\alpha})_{+}, k\in \mathbb{Z}_{+}^{odd}.
\end{equation}
Therefore, the Lax operator $ L_{c}$ and $R_{\alpha}$ have the dressing structures as follows
\begin{eqnarray*}
 L_{c}=\Psi_{c}D \d \Psi_{c}^{-1},\ \ \ R_{\alpha}=\Psi_{c}E_{\alpha}\Psi_{c}^{-1},
\end{eqnarray*}
where the operator $\d$ is equivalent to $a_1^{-1}\d_{t_{1}^{1}}+a_2^{-1}\d_{t_{1}^{2}}+...+a_\emph{N}^{-1}\d_{t_{1}^{\emph{N}}}$ and the dressing operator $\Psi_{c}$ should satisfy
\begin{equation}
\Psi_{c}^{*}=\Psi_{c}^{-1}.
\end{equation}
Here, eq.(\ref{h}) is the $C$ type condition of the $\emph{N}$-component CKP hierarchy.
In addition, the operators $A_{\alpha,k}$ satisfy the $C$ type condition as follows
\begin{equation}\label{i}
A_{\alpha,k}^{*}=-A_{\alpha,k},\ \ \end{equation}
and the dressing operator $\Psi_{c}$ satisfies the following Sato equation
\begin{equation}
\frac{\d \Psi_{c}}{\d t_{k}^{\alpha}}=-(L_{c}^{k}R_{\alpha})_{-}\Psi_{c}.
\end{equation}

\sectionnew{Quantum torus symmetry of the multi-component CKP hierarchy}
In order to construct the additional quantum torus symmetry of the multi-component CKP hierarchy, the operator $\Gamma_{c}$ and the Orlov-Schulman's operator $ M_{c}$ should be defined as
\begin{equation}
\Gamma_{m}=\sum\limits_{i\in \mathbb{Z}_{+}^{odd}}\sum\limits_{j=1}^{\emph{N}}it_{i}^{j}D^{-1}E_{jj}\d^{i-1},\ \ \ M_{c}=\Psi_{c}\Gamma_{m}\Psi_{c}^{-1}.
\end{equation}
The Lax equation of the multi-component CKP hierarchy can be described by the consistent conditions of the following set of linear partial differential equations
\begin{equation}
 L_{c}\omega_{c}(t;z)=z\omega_{c}(t;z), \frac{\d \omega_{c}}{\d t_{2n+1}^{j}}=( L_{c}^{2n+1}R_{j})_{+}\omega_{c}, \ \ t=(t_{1}, t_{3}, t_{5}, ...).
\end{equation}
Here the wave function can be defined as
\begin{equation}
\omega_{c}(t;z)=\Psi_{c}e^{\xi_{c}(t;z)},
\end{equation}
in which the wave operator of the multi-component CKP hierarchy is given by $\Psi_{c}=1+\sum\limits_{i=1}^{\infty}\omega_{i}\d^{-i}, i\in \mathbf{Z}$ and the the matrix function $\xi_{c}$ is defined as $\xi_{c}(t;z)=\sum\limits_{k\in \mathbb{Z}_{+}^{odd}}\sum\limits_{j=1}^{\emph{N}}t_{k}^{j}E_{jj}z^{k}$. In addition, it is easy to see $\d^{i}e^{xz}=z^{i}e^{xz}, i\in \mathbf{Z}$.\\
Basing on the above preparation, it is high time that we construct the additional quantum torus symmetry of the multi-component CKP hierarchy. The Lax operator $ L_{c}$ and the Orlov-Schulman's operator $ M_{c}$ satisfy the following matrix canonical relation
\begin{equation}
[ L_{c}, M_{c}]=1,
\end{equation}
in another, the Orlov-Schulman's operator $ M_{c}$ satisfies
\begin{equation}
 M_{c}\omega_{c}(z)=\d_{z}\omega_{c}(z),\ \ \frac{\d  M_{c}}{\d t_{k}^{j}}=[( L_{c}^{k}R_{j})_{+}, M_{c}],\ \ k\in \mathbb{Z}_{+}^{odd}.
\end{equation}
To any pair of integers $(m,n)$ with $m\geq0, n\geq0$, the matrix operator $A_{mnj}$ can be defined as
\begin{equation}
A_{mnj}= M_{c}^{m} L_{c}^{n}R_{j}-(-1)^{n}R_{j} L_{c}^{n} M_{c}^{m},
\end{equation}
and we have
\begin{equation}
\frac{\d A_{mnj}}{\d t_{k}^{s}}=[(L_{c}^{k}R_{s})_{+},A_{mnj}], k\in \mathbb{Z}_{+}^{odd}.
\end{equation}
Using the lemma (\ref{c}) and the eq.(\ref{i}), we have
\begin{eqnarray*}
A_{mnj}^{*}&=&( M_{c}^{m} L_{c}^{n}R_{j}-(-1)^{n}R_{j} L_{c}^{n} M_{c}^{m})^{*}\\
&=&(-1)^{n}R_{j} L_{c}^{n} M_{c}^{m}-(-1)^{2n} M_{c}^{m} L_{c}^{n}R_{j}\\
&=&(-1)^{n}R_{j} L_{c}^{n} M_{c}^{m}- M_{c}^{m} L_{c}^{n}R_{j}\\
&=&-( M_{c}^{m} L_{c}^{n}R_{j}-(-1)^{n}R_{j} L_{c}^{n} M_{c}^{m})\\
&=&-A_{mnj}.
\end{eqnarray*}
Therefore, it is easy for us to get that the $A_{mnj}$ satisfies the $C$ type condition
\begin{equation}\label{j}
A_{mnj}^{*}=-A_{mnj}.
\end{equation}
A matrix operator $C_{mnj}$ can be denote as
\begin{equation}
C_{mnj}:=e^{m M_{c}}q^{n L_{c}}R_{j}-R_{j}q^{n L_{c}}e^{m M_{c}},
\end{equation}
which further leads to
\begin{equation}\label{l}
C_{mnj}=\sum\limits_{p,s=0}^{\infty}\frac{m^p(n \log q)^{s}( M_{c}^{p} L_{c}^{s}R_{j}-(-1)^{s}R_{j} L_{c}^{s} M_{c}^{p})}{p!s!}=\sum\limits_{p,s=0}^{\infty}\frac{m^p(n \log q)^{s}A_{psj}}{p!s!}.
\end{equation}
By using eq.(\ref{j}), we will get the $C$ type anti-symmetry property of $C_{mnj}$ by the following calculation
\begin{eqnarray*}
C_{mnj}^{*}&=&(\sum\limits_{p,s=0}^{\infty}\frac{m^p(n \log q)^{s}A_{psj}}{p!s!})^{*}\\
&=&-\sum\limits_{p,s=0}^{\infty}\frac{m^p(n \log q)^{s}A_{psj}}{p!s!}\\
&=&-C_{mnj}.
\end{eqnarray*}
Therefore, the matrix operator $C_{mnj}$ satisfies the $C$ type condition
\begin{equation}
C_{mnj}^{*}=-C_{mnj}.
\end{equation}
The additional flows for the time variable $t_{m,n}^{j}$, $t_{m,n}^{*j}$ are defined on the basis of the quantum parameter $q$ as follows
\begin{equation}
\frac{\d \Psi_{c}}{\d t_{m,n}^{j}}=-(A_{mnj})_{-}\Psi_{c},\ \ \frac{\d \Psi_{c}}{\d t_{m,n}^{*j}}=-(C_{mnj})_{-}\Psi_{c},
\end{equation}
which are equivalent to
\begin{equation}
\frac{\d  L_{c}}{\d t_{m,n}^{j}}=-[(A_{mnj})_{-}, L_{c}],\ \ \frac{\d  M_{c}}{\d t_{m,n}^{*j}}=-[(C_{mnj})_{-}, M_{c}].
\end{equation}
In another, we can get
\begin{equation}
\d_{t_{l,k}^{*i}}(C_{mnj})=[-(C_{lki})_{-},C_{mnj}].
\end{equation}
With the reduction condition in the eq.(\ref{h}) on the generators of the additional torus symmetry flows, we construct the matrix operator $C_{mnj}$.
\begin{theorem}
The additional flows $\d_{t_{m,n}^{*j}}$ are symmetries of the multi-component CKP hierarchy, which commute with all flows $\d_{t_{k}^{j}}$ of the multi-component CKP hierarchy as
\begin{equation}
[\d_{t_{m,n}^{*j}},\d_{t_{k}^{j}}]=0.
\end{equation}
\end{theorem}
\textbf{Proof}. The proof is similar as the CKP hierarchy by using the Theorem \ref{k}. So the detail will be omitted here. Therefore, the additional flows $\d_{t_{m,n}^{*j}}$ of the multi-component CKP hierarchy commute with all flows $\d_{t_{k}^{j}}$ of the multi-component CKP hierarchy.\\
It is well known that the additional flows $\d_{t_{l,k}}$ of the CKP hierarchy form the $W_{\infty}$ algebra which has been shown in \cite{A}.
\begin{eqnarray*}
[\d_{t_{p,s}},\d_{t_{a,b}}] L_{c}=\sum\limits_{\alpha\beta}C_{\alpha\beta}^{(ps)(ab)}\d_{t_{\alpha\beta}} L_{c}.
\end{eqnarray*}
The additional flows $\d_{t_{l,k}^{s}}$ of the multi-component CKP hierarchy form the following $\emph{N}$-folds direct product of the $W_{\infty}$ algebra as follows
\begin{eqnarray*}
[\d_{t_{p,s}^{r}},\d_{t_{a,b}^{c}}] L_{c}=\delta_{rc}\sum\limits_{\alpha\beta}C_{\alpha\beta}^{(ps)(ab)}\d_{t_{\alpha,\beta}^{c}} L_{c}.
\end{eqnarray*}
Comparing to the quantum torus symmetry flows of the CKP hierarchy  given by
\begin{eqnarray*}
[\d_{t_{n,m}^{*}},\d_{t_{l,k}^{*}}]=(q^{ml}-q^{nk})\d_{t_{n+l,m+k}^{*}}, n, m, l, k \geq0,
\end{eqnarray*}
now it is high time that we construct the algebraic structure of the additional flows $\d_{t_{l,k}^{^{*j}}}$ of the multi-component CKP hierarchy.
\begin{theorem}
The additional flows $\d_{t_{l,k}^{*i}}$ of the multi-component CKP hierarchy form the $\bigotimes^{\emph{N}}QT_{+}$ algebra($\emph{N}$-folds direct product of the positive half of the quantum torus algebra QT), we get
\begin{equation}
[\d_{t_{n,m}^{*r}},\d_{t_{l,k}^{*j}}]=\delta_{rj}(q^{ml}-q^{nk})\d_{t_{n+l,m+k}^{*r}}, n, m, l, k\geq0,\ \ 1\leq r, j \leq \emph{N}.
\end{equation}
\end{theorem}
\textbf{Proof}. Similar to eq.(\ref{l}), we can rewrite the quantum torus flow which is a combination of the additional flows $\d_{t_{m,n}^{j}}$ as follows
\begin{eqnarray*}
[\d_{t_{n,m}^{*r}},\d_{t_{l,k}^{*j}}] L_{c}&=&[\sum\limits_{p,s=0}^{\infty}\frac{n^p(m \log q)^{s}}{p!s!}\d_{t_{p,s}^{r}},\sum\limits_{a,b=0}^{\infty}\frac{l^a(k \log q)^{b}}{a!b!}\d_{t_{a,b}^{j}}] L_{c}\\
&=&\sum\limits_{p,s=0}^{\infty}\sum\limits_{a,b=0}^{\infty}\frac{n^p(m \log q)^{s}}{p!s!}\frac{l^a(k \log q)^{b}}{a!b!}[\d_{t_{p,s}^{r}},\d_{t_{a,b}^{j}}] L_{c}\\
&=&\sum\limits_{p,s=0}^{\infty}\sum\limits_{a,b=0}^{\infty}\frac{n^p(m \log q)^{s}}{p!s!}\frac{l^a(k \log q)^{b}}{a!b!}\sum\limits_{\alpha\beta}C_{\alpha\beta}^{(ps)(ab)}\delta_{rj}\d_{t_{\alpha,\beta}^{r}} L_{c}\\
&=&(q^{ml}-q^{nk})\sum\limits_{\alpha,\beta=0}^{\infty}\frac{(n+l)^{\alpha}((m+k)\log q)^{\beta}}{\alpha!\beta!}\delta_{rj}\d_{t_{\alpha,\beta}^{r}} L_{c}\\
&=&(q^{ml}-q^{nk})\delta_{rj}\d_{t_{n+l,m+k}^{*r}} L_{c}.
\end{eqnarray*}
Therefore, we can derive a perfect quantum algebra of the multi-component CKP hierarchy which is similar to the quantum torus symmetry of the CKP hierarchy.

\sectionnew{Conclusions and Discussions}
In this paper, we construct the quantum torus symmetry of the CKP hierarchy and the multi-component CKP hierarchy. Meanwhile, the quantum torus symmetry flows form the quantum torus algebra which is given. In the further study, we are looking forward to giving out more quantum torus symmetry of the KP type integrable hierarchies and finding out the application of these quantum torus symmetries.

{\bf {Acknowledgements:}}
  This work is supported by the  National Natural Science Foundation of China under Grant No. 11571192  and K. C. Wong Magna Fund in
Ningbo University.


\begin{thebibliography}{AAA1}
\frenchspacing
\bibitem{B} L. A. Dickey, Soliton Equations and Hamiltonian Systems(2nd Edition) (World Scintific, Singapore, 2003).

\bibitem{C} A. Yu. Orlov, E. I. Schulman, Additional symmetries of integrable equations and conformal algebra reprensentation, Lett. Math. Phys. 12(1986), 171-179.

\bibitem{A} J. S. He, K. L. Tian, A. Forester, W. X. Ma, Additional Symmetries and String Equation of the CKP Hierarchy, Lett. Math. Phys. 81(2007), 119-134.

\bibitem{D} K. L. Tian, J. S. He, J. P. Cheng, Y. Cheng, Additional symmetries of constrained CKP and BKP hierarchies, SCIENCE CHINA Mathematics 54(2011), 257-268.

\bibitem{E} M. H. Li, C. Z. Li etal, Virasoro type algebraic structure hidden in the constrained discrete Kadomtsev-Petviashvili hierarchy, J. Math. Phys. 54(2013), 043512.

\bibitem{T} E. Date, M. Jimbo, M. Kashiwara and T. Miwa, Transformation groups for soliton equations III, J. Phys. Soc. Japan 50(1981), 3806-3812.

\bibitem{F} R. Dijkgraaf, E. Witten, Mean field theory, topological field theory, and multimatrix models, Nucl. Phys. B 342(1990), 486-522.

\bibitem{G} M. Douglas, Strings in less than one dimension and the generalized KdV hierarchy, Phys. Lett. B 238(1990), 176-180.

\bibitem{H} R. Block, On torsion-free abelian groups and Lie algebras, Proc. Amer. Math. Soc., 9(1958), 613-620.

\bibitem{I} J. M. Osborn, K. Zhao, Infinite-dimensional Lie algebras of generalized Block type, Proc. Amer. Math. Soc., 127(1999), 1641-1650.

\bibitem{J} X. Xu, Generalizations of Block algebras, Manuscripta Math., 100(1999), 489-518.

\bibitem{K} Y. Su, Quasifinite representations of a Lie algebra of Block type, J. Algebra, 276(2004), 117-128.

\bibitem{L} C. Z. Li, J. S. He, Y. C. Su, Block type symmetry of bigraded Toda hierarchy, J. Math. Phys. 53(2012), 013517.

\bibitem{M} C. Z. Li, J. S. He, Dispersionless bigraded Toda hierarchy and its additional symmetry, Reviews in Mathematical Physics, 24(2012), 1230003.

\bibitem{N} C. Z. Li, J. S. He, Block algebra in two-component BKP and D type Drinfeld-Sokolov hierarchies, J. Math. Phys. 54(2013), 113501.

\bibitem{O} C. Z. Li, J. S. He, Y. C. Su, Block (or Hamiltonian) Lie symmetry of dispersionless D type Drinfeld-Sokolov hierarchy, Commun. Theor. Phys. 61(2014), 431-435.

\bibitem{P} T. Nakatsu, K. Takasaki, Melting Crystal, Quantum Torus and Toda Hierarchy, Commun. Math. Phys. 285(2009), 445-468.

\bibitem{Q} C. Z. Li, J. S. He, Quantum Torus symmetry of the KP, KdV and BKP hierarchies, Lett. Math. Phys. 104(2014), 1407-1423.
\bibitem{S} M. H. Tu, On the BKP hierarchy: additional symmetries, Fay identity and Adler-Shiota-van Moerbeke formula, Lett. Math. Phys. 81(2007), 93-105.

\bibitem{R} C. Z. Li, J. S. He, Supersymmetric BKP systems and their symmetries, Nuclear Physics B 896(2015), 716-737.



\end{thebibliography}
\end{document}